# Overtures to the pulsational instability of ZZ Ceti variables

**Alfred Gautschy[1,2], Hans-Günter Ludwig[2], Bernd Freytag[3]**

[1] Astronomisches Institut der Universität Basel, Venusstr. 7, CH-4102 Binningen, Switzerland

[2] Max-Planck-Institut für Astrophysik, Karl-Schwarzschild Str. 1, D-85740 Garching, Germany

[3] Institut für Astronomie und Astrophysik der Universität, D-24098 Kiel, Germany

**Abstract.** Results of nonradial, nonadiabatic pulsation calculations on hydrogen-rich white dwarf models are presented. In contrast to earlier attempts, the modeling builds on hydrodynamically simulated convective surface layers supplemented with standard interior models. Based on our stellar models and despite of various simple attempts to couple convection and pulsation we could not reproduce theoretically the presently adopted location of the observed blue edge of the ZZ Ceti variables. When the convective efficiency is high enough we found a sensitive dependence of the stability properties of the $g$ modes on the pulsational treatment of shear within the convection zone.

## 1. Introduction

When white dwarf stars evolve across the Hertzsprung-Russell (HR) diagram they encounter episodes of oscillatory instabilities. In the major families of white dwarfs – DA, DB, and DO – partial ionization of the respective dominant chemical constituents in the envelopes are believed to destabilize numerous gravity modes. The different envelope structures and the different ionization potentials involved in the DA, DB, and DO variabilities place the instability regions at different locations along the white-dwarf cooling tracks.

The frequently observed rich spectra of pulsation modes of oscillating white dwarfs made them predestined targets for asteroseismology. Indeed, unprecedented accuracy in the determination of stellar parameters (mass, chemical stratification, rotation rate, or magnetic field strength) is being claimed (Winget et al. 1994, Bergeron et al. 1993, for example). The DAV (ZZ Ceti) stars have the smallest number of unstable oscillation modes and therewith the simplest powerspectra of all variable white dwarfs. The seismic analyses rely on the observed oscillation frequencies and their separations so that the accompanying theoretical modeling needs to provide accurate adiabatic eigenspectra only (see e.g. Bradley & Winget 1991, Brassard et al. 1992). At the level of the presently achievable

*Send offprint requests to:* A. Gautschy, at address 1

accuracy in seismic studies, nonadiabatic effects on the oscillation frequencies can be neglected. Nonadiabatic processes determine, on the other hand, the exact locations and extents of the instability regions.

The existence of nonradial oscillations in ZZ Ceti stars – the targets of this contribution – is attributed to the $\kappa$-mechanism caused by partial hydrogen (H) ionization (Dolez and Vauclair 1981, Winget et al. 1982). The H-ionization region of these stars is convectively unstable and a *major* part of the energy can be transported by material motion. As the dominant pulsational driving was found to occur at the base of this convection zone the particulars of treating the convection in the stellar structure calculation and in the perturbation equations are important. From the very beginning of linear nonadiabatic oscillation analyses of ZZ Ceti models the undetermined role played by convection in the destabilization of $g$ modes lessened the credibility of the results.

All investigators having worked on locating the DAV instability region (e.g. Bradley & Winget 1994, Fontaine et al. 1994, and references therein to previous work) agreed on the efficiency of convection to play a crucial role for the final outcome. Depending on the choice of the convective efficiency, usually by adopting a particular dialect of the mixing-length theory (MLT), the blue edge of the ZZ Ceti instability region can easily shift by more than $1\,000\,\mathrm{K}$ in effective temperature. Hitherto, the canonical approach was to adjust the convective efficiency, i.e. the depth of the convection zone, such that the location of the theoretical blue edge agreed with the effective temperatures of the hottest observed DAV stars.

The spectroscopic calibration of the physical parameters of DAV stars is rather involved. The analyses of observed white dwarf spectra still tolerate a considerable spread in $\log g$ and in $T_{\mathrm{eff}}$. The results depend sensitively on highest-quality spectra and also on the physics implemented in the calculations of the model spectra. The star G117-B15A was considered to be among the hottest ZZ Ceti variables for a long time. Recently, G117-B15A was recalibrated by Koester et al. (1994) and its effective temperature was lowered from $13\,000\,\mathrm{K}$ (Weidemann & Koester 1984) to about $12\,250\,\mathrm{K}$. In another recent publication (Bergeron et al. 1995), the same star was attributed an even lower effective temperature of $11\,620\,\mathrm{K}$ (cf. Fig. 1). Based on the distribution of pulsation amplitudes in different wavelength bands, Robinson et al. (1995) published also a recalibration of G117-B15A. They found a correlation between $\log g$ and $T_{\mathrm{eff}}$ that allowed a range of different effective temperatures, at dif-ferent surface gravities, to fit the observational data (see the dash-dotted line in Fig. 1). A particular $T_{eff}$ – at 12 375 K – was eventually adopted with the help of a recent spectroscopic determination of $\log g$.

The hottest member – G226-29 – in the ZZ Ceti sample of the Bergeron et al. (1995) study has $T_{eff} = 12\,460$ K, $\log g = 8.29$ and lies supposedly close to the blue edge of the instability domain. From the distribution of the variables on the $T_{eff}$ – $\log g$ plane it is conceivable that the location of the blue edge is not vertical. From Fig. 1 we estimate the blue edge to lie at about 12 000 K for $\log g = 8.0$. The magnitude of disagreement in assigning effective temperatures and surface gravities to observed ZZ Ceti variables is remarkable and should be kept in mind when comparing theory and observation of white dwarf oscillations. Nevertheless, the new data hint at a cooler blue edge than was assumed hitherto.

Even if the temperature shift of the blue edge is considered to be of secondary importance for the DAV phenomenon, the role played by convection, in particular its efficiency, is decisive for $g$-mode instability and it remains to be understood. The theoretical studies of ZZ Ceti pulsations depend on the convection in a two-fold way. First, the interior models need particular choices of mixing-length parameters to allow unstable $g$ modes to occur at effective temperatures and surface gravities that are compatible with observations. Second, the oscillation equations themselves contain terms where the energetic contribution, i.e. the time-dependence of the convective flux, needs to be specified. Stability or instability might be dominated by the form of the interaction formalism. Previous studies usually assumed a frozen-in flux approximation to describe this interaction. Interpretations of such calculations led to the concept of "convective blocking" to describe the large positive contributions to the work integral at the base of the hydrogen convection zone (Cox et al. 1987, Pesnell 1987, Brickhill 1991). The physical significance of the mechanism of convective blocking is not fully understood or accepted yet (see also Brickhill 1991). The use of the frozen-in flux approximation in modeling ZZ Ceti stars is based more on feasibility rather than on physical adequacy, hence it is an open problem if something like convective blocking is really at work in these stars.

The following study intends to contribute towards an understanding of the influence of the convection zone in the background models and the role it plays in the oscillation equations for stability analyses of ZZ Ceti variables. We present results from our attempt to circumvent the problem of ad hoc tuning the convective efficiency in a simplified convection theory by using suitably post-processed data from two-dimensional hydrodynamical simulations of the surface regions of white dwarfs (Ludwig et al. 1994, Freytag et al. 1995). The processed hydrodynamic surface layers were fused consistently to interior models in hydrostatic and thermal equilibrium. Linear, nonradial, nonadiabatic stability properties were investigated on these hybrid models. Methodical aspects of this approach are presented in Sect. 2 and in the Appendix. Section 2 additionally reviews properties of the hydrodynamically simulated convection zones. Results of the stability analyses and of our simple attempts to couple convection with the oscillation equations are described in Sect. 3. Section 4 contains a critical discussion. Section 5 closes the paper with the conclusions and suggestions for the origin of the encountered discrepancies.

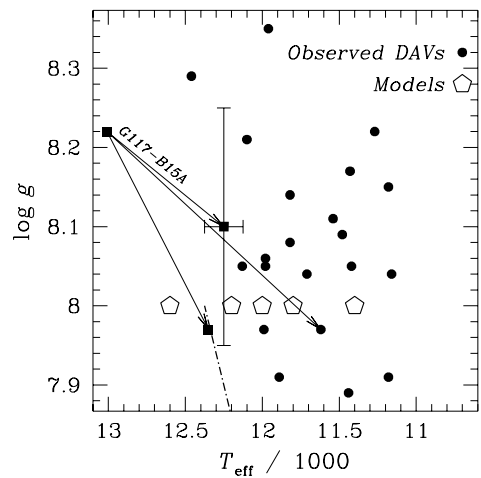

**Fig. 1.** Distribution of ZZ Ceti variables in the $T_{eff}$ – $\log g$ diagram. Dots represent the the latest calibrations according to Bergeron et al. (1995). The relocation experienced by G117-B15A from the Weidemann & Koester (1984) calibration to the positions determined according to Koester et al. (1994) (solid square with error bars), Bergeron et al. (1995) (with the lowest $T_{eff}$), and Robinson et al. (1995) is indicated. The open pentagons indicate the positions of our stellar models.

## 2. Modeling DA white dwarfs and their oscillations

This section, together with Appendix A that is devoted to technical details, describes the methodical tools used to calculate equilibrium structures of the white dwarf models and the numerical treatment of the nonradial oscillation problem.

### 2.1. The hydrodynamical envelope models

Hydrodynamical simulations have been performed to derive the structure of the convective surface layers of our white dwarf models. In these calculations the basic hydrodynamical equations are solved together with the equation of radiative transfer for a homogeneous, compressible, viscous, and stratified medium in two-dimensional (2D) Cartesian geometry. A realistic equation of state and realistic opacities including the Lyman-$\alpha$ satellite features (cf. Allard et al. 1994) are used. The transfer equation is solved in LTE approximation accounting for the non-local and frequency dependent character of the radiation field. In the present investigation usually 7 frequency points are used in the calculation of the radiative energy transport. In particular, the effects of the localized opacity enhancements from the Lyman-$\alpha$ features around 1400 Å and 1600 Å are represented adequately. This – in a hydrodynamical context – rather detailed treatment of the frequency dependence of the radiation field has turned out to be necessary for a successful modeling of the photospheric temperature stratification. A detailed description of the program code and results of white-dwarf simulations can be found in Ludwig et al. (1994) and Freytag et al. (1995).

In the simulations, the temporal evolution of the convective flow was followed over a sufficiently long time to ensure that a dynamically and thermally relaxed state was reached. An one-dimensional (1D) envelope structure was derived by averaging the physical quantities temporally and spatially on surfaces of constant geometrical depth. The time interval for this averaging is sufficiently long that statistically stable mean values were obtained. From the resulting data the mean gas pressure $\langle P \rangle(z)$ and temperature $\langle T \rangle(z)$ profiles were taken as basic input variables for the subsequent construction of the stellar model and the stability analysis. Secondary input parameters (e.g. sound speed) — especially for the stability analysis — are calculated from $\langle P \rangle$ and $\langle T \rangle$. It should be noted that we might not have used the optimal procedure to transfer the properties important for pulsations, i.e. quantities determining the propagation of gravity waves, from a horizontally inhomogeneous stratification to an 1D representation. It is conceivable that averaging certain quantities on surfaces of constant *acoustical* depth provides a closer fit to the inhomogeneous situation. Nevertheless, we believe that our approach covers the main effects introduced by the hydrodynamical models so that we have not investigated this question any further.

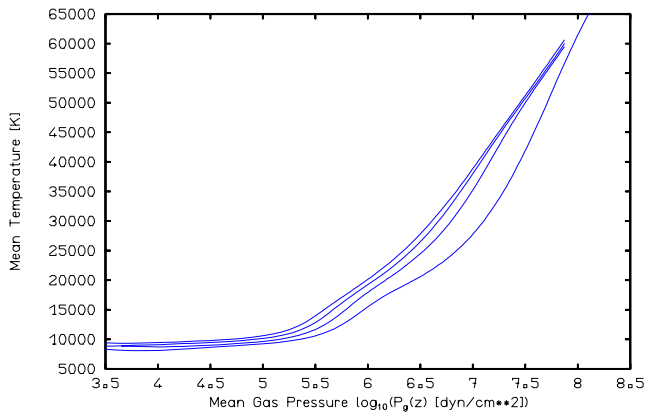

**Fig. 2.** Mean temperature versus mean pressure for the hydrodynamical envelope models with effective temperatures increasing from $11\,400\,\mathrm{K}$ (bottom line) in steps of $400\,\mathrm{K}$ (Models A, B, D, E, cf. Tab. 1).

Figures 2, 3, and 4 show mean temperature, convective heat (enthalpy) flux, and entropy as a function of the mean pressure for our set of envelope models. Five surface models were considered; to demonstrate the systematics more clearly only four models (A, B, D, and E, cf. Tab. 1) with equidistant spacing in $T_{\mathrm{eff}}$ are displayed. The sequence is not completely homogeneous; in the coolest model the radiative transfer is treated using 1 frequency point (gray approximation) instead of 7 as in the hotter ones. The computational boxes had different geometric sizes in the different models depending on the depth of the convective layers. Model A extends deepest; the photosphere of model B allows only for a less extended photosphere leading to the artificial increase of the entropy near the upper boundary that is visible in Fig. 4. Figure 3 shows how the fraction of the convective heat flux increases with decreasing effective temper-

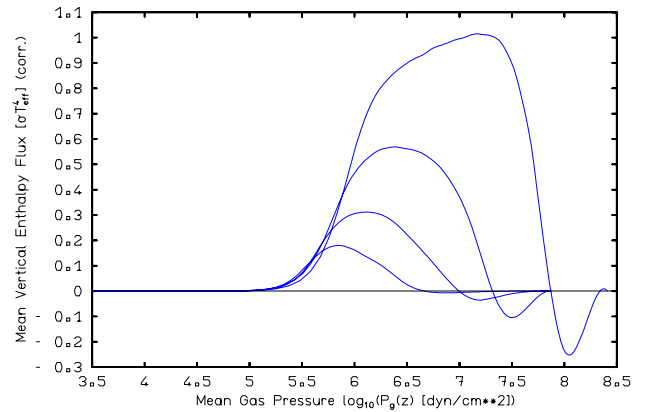

**Fig. 3.** Mean enthalpy flux versus mean pressure for hydrodynamical envelope models (cf. Fig. 2). The maximum flux increases monotonically from the hottest to the coolest model.

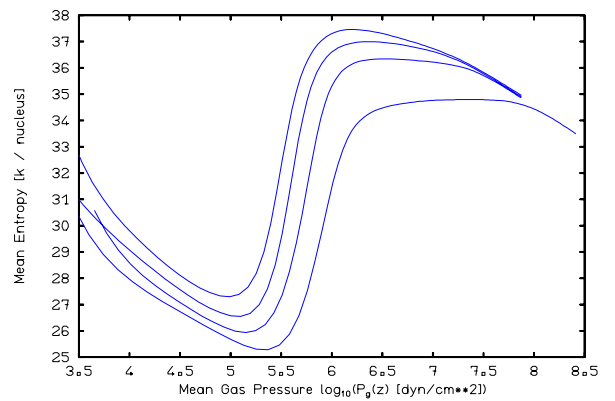

**Fig. 4.** Mean entropy versus mean pressure for hydrodynamical envelope models (cf. Fig. 2). The minimum entropy increases monotonically from the coolest to the hottest model.

ature. Apart from the uncertainty of the quantitative choice for the mixing-length parameter such behavior is also encountered in models based on MLT. Standard MLT does not include the effects of overshooting that leads to the regions with negative heat flux in the deeper layers. A close inspection of model A reveals a maximum heat flux that supersedes the nominal total flux by a small amount. Again, unlike in MLT approaches, in the hydrodynamical models the convective motions are accompanied by a significant flux of kinetic energy which is directed downward and balances the flux budget. Finally, by comparing Fig. 3 with Fig. 4 we realize that — particularly in the hotter models — subadiabatic regions where $\mathrm{d}\langle S \rangle/\mathrm{d}z > 0$ do not match exactly domains with positive convective flux $F_{\mathrm{h}} > 0$. This so called "entropy gradient reversal" phenomenon (cf. Skaley & Stix 1991, Chan & Gigas 1992, Ludwig et al. 1994) is a typical feature arising from a non-local description of convection. Clearly, this affects the propagation properties of gravity waves in the convection zone.

### 2.2. The interior models

The interior models, i.e. complete equilibrium solutions attached to the bottom of hydrodynamic envelopes, of the white dwarf models were computed with a shooting code. Some technical details of the method are presented in the Appendix. The model stars were chemically stratified with an a priori fixed parameterization of the thickness of the hydrogen, helium, and carbon/oxygen layers. Also the abundance profiles describing the transition from the H to He and from the He to C/O zones were functionally prescribed. The surface hydrogen-layer thickness varied between $10^{-4}$ and $10^{-7}M_\star$. The adjacent pure helium shell extended always to a depth of $0.98M_\star$. The remaining core consisted of a mixture of carbon and oxygen with C/O = 1/4 in mass and a remaining heavy-element abundance of $9.6 \times 10^{-3}$. A representative case of the chemical stratification is shown in Fig. 5. The model has a hydrogen surface layer thickness of $\Delta M_H = 1 \times 10^{-4}M_\star$. The quantity $q_H$ stands for the relative mass depth of the hydrogen layer: $q_H \equiv \Delta M_H / M_\star$. The form of the parameterized transition layers compares satisfactorily with self-consistently calculated ones in diffusive equilibrium. The averaged hydrodynamical layers contain between 80 and 110 gridpoints. The interior models are constructed with typically about 800 gridpoints.

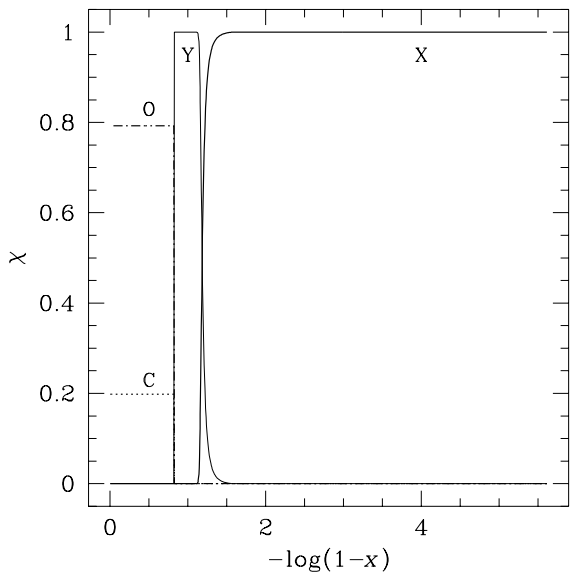

**Fig. 5.** Representative example, model B4, of the relative mass fractions of chemical abundances $\chi$ in the stellar models. The depth in the star is expressed as a function of the relative radius $x$. The labels at the different curves stand for hydrogen (X), helium (Y), carbon (C), and oxygen (O). The transition depths and the transition structure from the C/O core to the He shell is the same in all models; merely the mass of the hydrogen envelope was varied.

Table 1 lists the global parameters for a subset of DA stellar models that were obtained with our iterative shooting method; they served as background models on which the stability anal-yses were performed. In the following we refer to them as the "HD models". Note that luminosity and mass of the relaxed models are free parameters and must be iterated to obtain a hydrostatic model in thermal equilibrium. The distribution of the model stars on the $T_{\text{eff}} - \log g$ plane and their relation to recent observational data can be inferred from Fig. 1.

**Table 1.** Global parameters of the white dwarf models. For all models the acceleration of gravity is $\log g = 8.0$ at the surface.

| ID | $T_{\text{eff}}/\text{K}$ | $\log q_H$ | $M_\star/M_\odot$ | $\log L_\star/L_\odot$ | $\log R_\star/R_\odot$ |
|---|---|---|---|---|---|
| A4 | 11 400 | -4.0 | 0.6109 | -2.594 | -1.8881 |
| A7 | 11 400 | -7.0 | 0.5960 | -2.604 | -1.8935 |
| B4 | 11 800 | -4.0 | 0.6190 | -2.533 | -1.8878 |
| B7 | 11 800 | -7.0 | 0.5969 | -2.644 | -1.8932 |
| C4 | 12 000 | -4.0 | 0.6124 | -2.504 | -1.8876 |
| C7 | 12 000 | -7.0 | 0.5973 | -2.515 | -1.8930 |
| D4 | 12 200 | -4.0 | 0.6129 | -2.475 | -1.8875 |
| E4 | 12 600 | -4.0 | 0.6138 | -2.420 | -1.8871 |

Mostly we calculated the convective flux directly from averaged hydrodynamical simulation data. As the kinetic flux contributed only little, the convective flux was always computed from the enthalpy flux. We used the ratio of the mean enthalpy flux (see Fig. 3) to the total flux to express the fractional energy transport by convection. The values obtained therefrom can differ from those deduced from $1 - \nabla_0/\nabla_{\text{rad}}$ ($\nabla_0$ denotes the actual temperature gradient of the stellar envelope). A significant difference was found only in the 11 400 K model, however; the maximum relative convective flux resulting from the enthalpy flux is 0.996 compared with 0.90 calculated with $\nabla_0$ and $\nabla_{\text{rad}}$.

### 2.3. The nonradial oscillation treatment

Nonadiabatic, nonradial oscillation modes were calculated with the Riccati method in a version closely related to the one described in Glatzel & Gautschy (1992). For the applications in this paper we accounted for the advection terms in the Eulerian perturbation quantities in chemically inhomogeneous regions of the star to guarantee $\delta\chi/\chi \equiv 0$, i.e. vanishing Lagrangian perturbation of the abundance of any atomic species $\chi$ throughout the stellar models. (In the following we will use $\delta$ to denote the Lagrangian perturbation and a dash for the Eulerian perturbation.) In all cases, we computed the complex eigenvalues $(\sigma_R, \sigma_I)$ for the full sixth-order system of equations, i.e. including the perturbation of the gravitational potential. Eigenvalues and characteristic frequencies are being expressed in units of $\sqrt{3GM_\star/R_\star^3}$ which amounts to about 0.6 sec$^{-1}$ in our cases.

The DA white dwarfs are examples of stars with extreme ratios of the thermal to the dynamical time-scale. Too large values of this ratio can give rise to numerical instabilities in the inversion of the matrix arising in relaxation techniques. Lee & Bradley (1993) suggested an hybrid method to deal with the nonradial, nonadiabatic oscillation problem in a numerically stable way. The numerical properties of the Riccati approach and the associated shooting method do not depend on possibly extreme time-scale ratios. We were always able to use the same set of equations throughout the whole star without encountering numerical difficulties. The required accuracy of the eigenvalues enforced the distribution and the total number of the gridpoints chosen by our self-adaptive initial-value integrator. Typically, of the order of 3000 gridpoints were used for an eigensolution. Very high-order overtones could contain up to 10 000 gridpoints. The physical quantities of the background stellar models are always interpolated with the monotonized cubic method described in context of the calculation of the Brunt-Väisälä frequency.

The inner boundary conditions were formulated in the stellar center where the requirement of regularity of the perturbed physical quantities can be used to derive the necessary algebraic equations. The outer boundary was chosen at an optical depth of $\tau = 0.025$. For consistency with other results in the literature, we implemented the *physical* conditions suggested by Osaki & Hansen (1973).

Since the calculation of the Brunt-Väisälä frequency caused some debates in recent years, we shortly mention our approach – which differs from the ones presently referred to in the literature. We performed monotonized cubic interpolations of the density stratification with respect to pressure and calculated derivatives directly from the interpolating polynomial. As long as the background-model data was sufficiently smooth (i.e. without numerical noise) the interpolation method suggested by Steffen (1990) proofed to be very efficient and robust. An example of the high-quality derivatives that were obtained on our stellar models can be seen in the Brunt-Väisälä frequency of Fig. 6 for model B4. We emphasize that the steep He – C/O transition (at $\log(1 - q) = -2$ in Fig.6) is numerically well resolved. In a few cases, we had the opportunity to compare our numerical approaches with those of P. Bradley (see further down in the paper). The agreement of the Brunt-Väisälä frequency was very good except for small deviations in the compositional transition layers. The magnitudes of the disagreements were insignificant for the $g$-mode cavities.

## 3. Results

The convective turn-over time in the superficial convection zone of the DAV stars is much smaller than the oscillation time-scale ($\tau_{osc}/\tau_{conv} \gtrsim 1000$). The thermal time-scale of deep convection zones can become comparable with the oscillation periods sought and, additionally, most of the energy flux is then carried by material motion. This might be the reason for our inadequate understanding of the role the surface convection zone plays in driving the DAV oscillations. In a first attempt, we studied the stability properties of low-$\ell$ modes with the canonical assumption of "frozen-in" convective flux. In our formulation of the oscillation equations this simplification is easiest implemented as $\mathbf{F}'_C = 0$. In the literature, the most frequently encountered formulation is $\delta \nabla \cdot \mathbf{F}_C = 0$ or $\delta(\nabla \cdot \mathbf{F}_C/\rho) = 0$. Nevertheless, the correspondence between the two implementations should be close enough to serve as a first guideline.

Figure 7 shows representative results for low-degree, low-order modes of model B7. Reasonably expressed trapping cycles can be seen in the period separations, $\Delta\Pi$, displayed in the middle panel. The period-separation minima concur mostly with local maxima in the imaginary parts of the eigenfrequencies. Such a behavior agrees with the canonical picture of trapped modes having smaller kinetic energy than non-trapped modes. In the weakly nonadiabatic limit $\sigma_I/\sigma_R \approx$ $-W(R_\star)/4\pi E_{kin}$ obtains. Hence, a reduced kinetic energy of a mode translates into a more efficient damping or driving, depending on the sign determined by $W(R_\star)$, the total work done over a cycle by the particular oscillation mode; the meaning of the symbols is adapted from Saio & Cox (1980). The dependence of $E_{kin}$ on period, supporting the above picture, is shown in the top panel of Fig. 7.

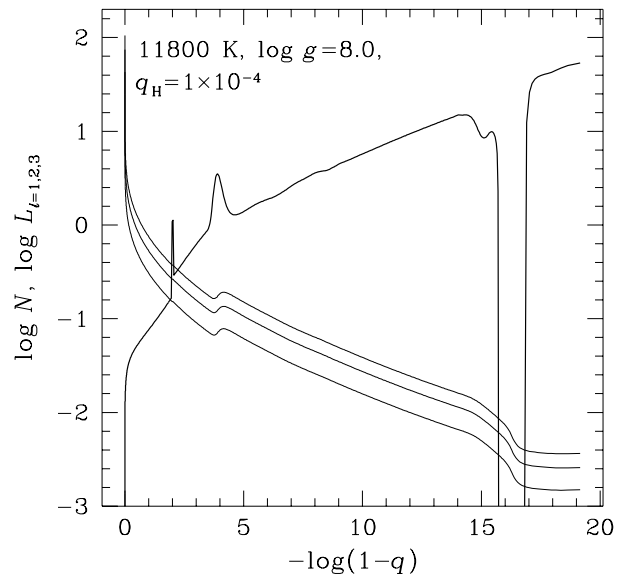

**Fig. 6.** Brunt-Väisälä and Lamb frequencies (for $\ell = 1, 2, 3$) for model B4. The functional behavior of both characteristic quantities is representative for all the models studied.

The marginal instabilities of one of the radial overtones appearing in all studied spherical degrees ($\ell = 1, 2,$ and 3) in Fig. 7 are not connected with destabilizing agents in the *surface* layers. The kinetic-energy density of each of the unstable modes has significant amplitude only in the C/O core and partially in the He shell of the models. The driving occurs around the sharp transition from the C/O core to the He shell. Figure 6 shows the narrow local bumps of the Brunt-Väisälä frequency at the C/O – He ($-\log(1 - q) = 2$) and at the He – H ($-\log(1 - q) = 4$) transition. For suitable frequencies such features can confine oscillation modes to sub-cavities within the total propagation region. We remind that our transition structures are analytically prescribed in the stellar models so that the trapping properties depend on our choice of the location and the steepness of the transition profiles. It is indeed the trapping of the (unstable) modes in the centermost regions that is responsible for the absence of a significant dependence of the driving rate on effective temperature and hydrogen-layer mass in the model sequence. The lowest panel in Fig. 12 shows the typical behavior of the logarithmic derivative of the opacity with respect to pressure at constant temperature $\kappa_T$ that is similar in all HD models. The deep interior, made up of He and of the CO core, is characterized by positive $\kappa_T$ that destabilizes a mode under a favorable behavior of the eigenfunction. From Fig. 7 we further learn that the imaginary parts of the eigenfrequency of fixed radial-order modes grow systematically when the spherical degree increases; simultaneously the oscillation periods shorten.

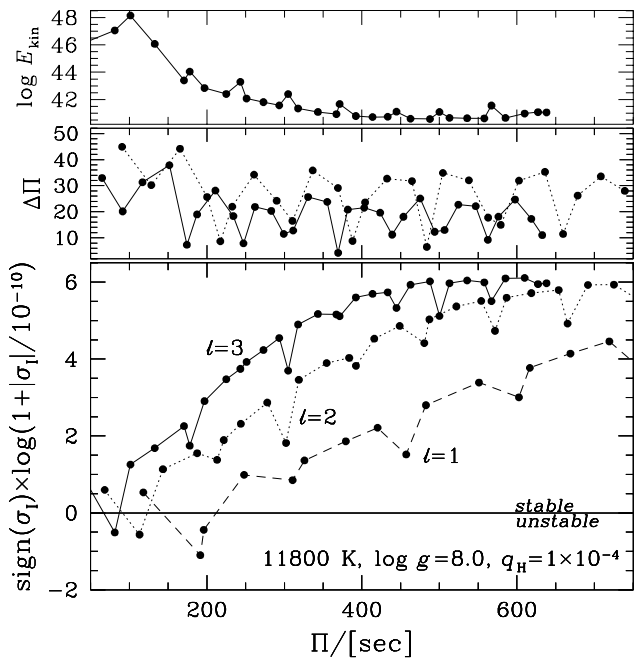

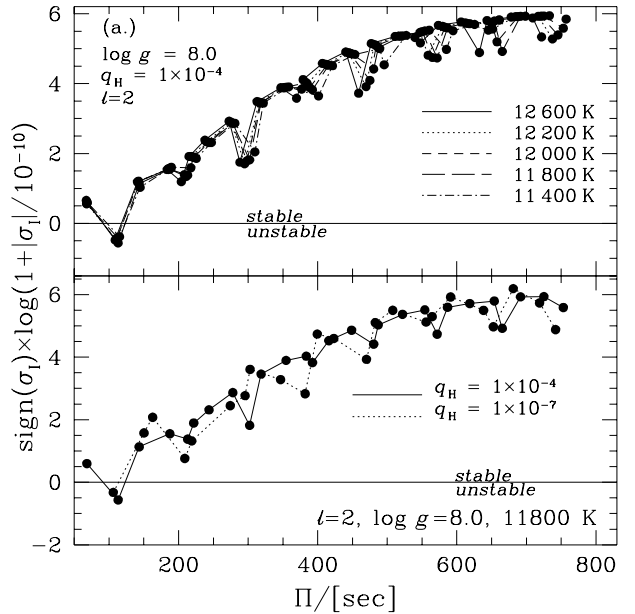

**Fig. 7.** The top panel displays the run of the kinetic energy of the $\ell = 3$ modes as a function of their periods. The middle panel shows the period spacing of the nonadiabatic periods for $\ell = 2$ and 3 modes of model B4. The lower panel shows the run of the damping rates of some of the lower-order modes of degree one to three. The scaling of its ordinate is intended to conveniently represent the huge dynamical range of the imaginary parts of the eigenvalues.

**Fig. 8. a** Run of the imaginary parts of the eigenfrequencies for different effective temperatures at the same spherical degree and the same hydrogen mass depth. **b** Effect on the imaginary part of the eigenfrequencies of varying the hydrogen mass-depth for $\ell = 2$ modes in 11 800 K models. In both panels, frozen-in flux treatment was adopted for convection.

We want to emphasize that the trapping properties encountered in our models with thick hydrogen layers differ qualitatively from those described previously (e.g. Brassard et al. 1992). Generally, the modes reside in the stellar envelope without being particularly sensitive to the compositional transition structures. A few modes with appropriate eigenfrequencies distinguish themselves by their confinement to the deep interior. This leads to an enhanced kinetic energy (see top panel of Fig. 7), a local minimum in the period spacing (middle panel), and a reduced imaginary part of the eigenfrequency (bottom panel).

Untrapped modes (of equal spherical degree) tend to increase their damping rates towards higher effective temperatures. The trapped modes show the opposite trend: trapped modes in hotter models are less damped than trapped modes in cool models. The global run of the damping versus period is the same at all effective temperatures considered (see Fig. 8 a). Panel b in Fig. 8 demonstrates the marginal effect on the excitation/damping rates of oscillation modes upon changing the envelope hydrogen mass of the HD models. In none of the sequences studied did we encounter a case where changing the magnitude of the hydrogen envelope mass changed the stability properties of the $g$ modes.

We close this section by stressing again that with the approximation of frozen-in convective flux we found *no* evidence for ZZ Ceti – like $g$-mode instabilities for any kind of parameter settings in our model sequence based on hydrodynamic convective envelopes.

### 3.1. Tests on models with thick convection zones

To ensure the integrity of the stellar-oscillation computer code we had access to three stellar models from a white dwarf evolutionary sequence (at 12 440, 12 080, and 11 720 K) calculated and kindly made available by P. Bradley (1995, private communication). All models had thick convection zones resulting from a ML3 mixing-length prescription (Bradley & Winget 1994). We tried to recover unstable oscillation modes that were claimed to be present in the cooler two models. The calculations were restricted to the lowest three nonradial degrees. Applying the previously mentioned approximation for the convection treatment: $\mathbf{F}'_C = 0$ yielded no unstable modes, but a few only marginally stable ones, at 12 440 K. In the lower-temperature models, a rich spectrum of unstable modes was encountered for spherical degrees $\ell = 1, 2$, and 3; 28 to 35 modes were pulsationally unstable with periods between 200 and 2000 sec. A sample of results is shown in Fig. 9; it contains the lowest-order $\ell = 3$ modes for the 12 080 K (No.98 in Bradley's series) model. The frozen-in results are shown as the points connected with the solid line. The unstable modes extend to very high overtones, only above about 1000 sec does enhanced dissipation within the convection zone of partial H-ionization (which is a region with $\mathrm{d}\kappa_T/\mathrm{d}r > 0$) win over adjacent driving in particular at the base of the convection zone. Neglecting convective energy transport completely in the nonradial perturbation equations resulted in *no* unstable modes at all. This emphasizes again the well-known necessity of a physically realistic treatment of the convective-flux contribution in the perturbation equations when dealing with ZZ Ceti variables. The numerical experiments on the model stars of Bradley proved unambiguously that our numerics is able to reliably detect unstable oscillation modes if they exist.

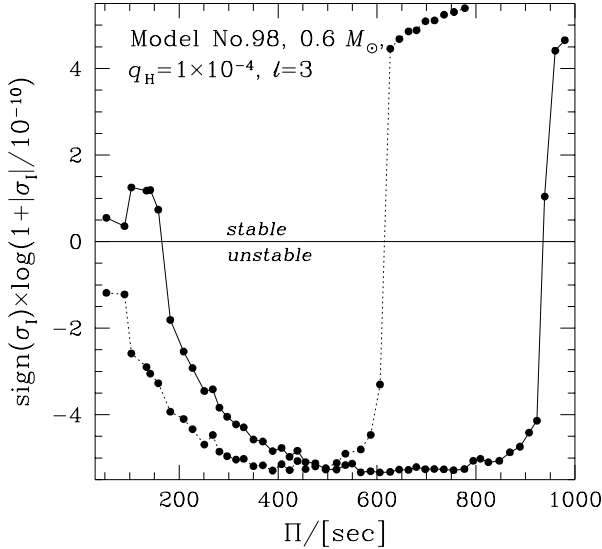

**Fig. 9.** Variation of the damping rates as a function of period for $\ell = 3$ modes in the $12\,080\,\mathrm{K}$ model of Bradley. The full line connects the results from the frozen-in flux approach; the dotted line is based on the outcome from instantaneous adaptation and vanishing shear within the convection zone.

### 3.2. Modified convection coupling

The turn-over times of convective eddies in our models are much shorter than the typical periods in which we are interested based on observational evidence. Instantaneous adaptation of convection to its oscillating environment appears to be a more realistic approximation. Instead of demanding $(\nabla \cdot \mathbf{F}_C')_r = 0$, no fluctuation of the radial component of the Eulerian perturbation of the convective flux, we let it adapt to the local perturbations. The resulting *perturbation* equation is rather lengthy so that we do not reproduce it but mention only the original equation that was perturbed and the assumptions having entered. We write for the convective flux

$$(\mathbf{F}_C)_r = \left\{ \frac{16\sigma_{\mathrm{SB}}}{3} \left(\nabla_{\mathrm{rad}} - \nabla_0\right) \right\} \frac{T^4 V}{\rho \kappa r}. \qquad (1)$$

The quantity $\sigma_{\mathrm{SB}}$ stands for the Stefan-Boltzmann constant and $V \equiv -\mathrm{d}\log P/\mathrm{d}\log r$, i.e. one of the homology invariants. The remaining symbols have their usual meanings. For simplicity we neglected the perturbation of the terms in braces on the right-hand side. In particular, neglecting the perturbation of $\nabla_{\mathrm{rad}} - \nabla_0$ means that the size of the convection zone is kept fixed in mass during the oscillation cycle. We refer to this approximation as IA1. In the energy equation, we implemented the Eulerian perturbation of the above equation to compute the component $(\nabla \cdot \mathbf{F}_C)_r$. Due to persisting ignorance, we assumed a vanishing contribution from the tangential (to spherical shells) components of the divergence: $(\nabla \cdot \mathbf{F}_C')_\perp \equiv 0$. We estimate this to be not too severe a limitation as only low spherical degrees were considered.

Upon repeating the stability analyses with the assumption of IA1 instantaneous adaptation of convection we did find no pulsational instabilities in the HD model sequence. The damping rates even changed marginally only when changing the convective adaptation picture. In the Bradley models with their deep, efficient convection zones, the mode spectrum (periods and damping/excitation rates) did not react on the change from frozen-in flux to instantaneous adaptation. In particular, the unstable mode spectrum remained unchanged.

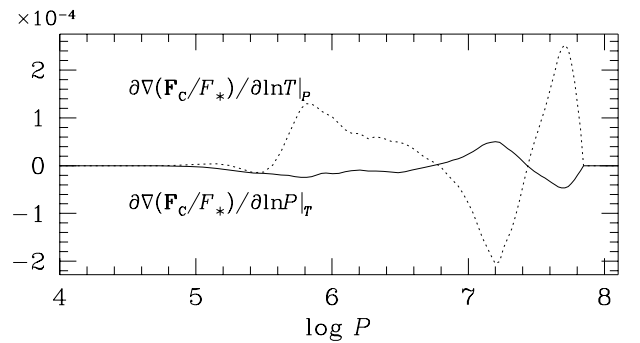

**Fig. 10.** Spatial run of the expansion coefficients $(\partial \nabla \cdot \mathbf{F}_C/\partial \ln P)_T$ and $(\partial \nabla \cdot \mathbf{F}_C/\partial \ln T)_P$ to express the the divergence of the perturbed convective flux. The numerical values were computed from the HD envelopes simulations.

For the $12\,000\,\mathrm{K}$ model we attempted a more realistic prescription of instantaneous adaptation of the convective flux (IA2). We approximated the perturbation of the divergence of the convective flux perturbation which appears in the energy equation by

$$\nabla \cdot \mathbf{F}_C' \approx \left(\frac{\partial \nabla \cdot \mathbf{F}_C}{\partial \ln P}\right)_T \cdot \frac{P'}{P} + \left(\frac{\partial \nabla \cdot \mathbf{F}_C}{\partial \ln T}\right)_T \cdot \frac{T'}{T}. \qquad (2)$$

The spatial form of the expansion coefficients displayed in Fig. 10 was derived from the HD envelopes utilizing the following idea: due to the small thermal and dynamical time-scales the convectively active layers stay close to states of hydrostatic and thermal equilibrium during the pulsation cycle. These states can be represented by a sequence of background models of different effective temperatures. A certain mass element experiences changes of $P$ and $T$ according to its various depths in the quasistatic stratifications. The convective flux at these depths is known from the background models. The changes of $P$ and $T$ are not independent but related by the constraint that the surface layers pass through quasistatic configurations. Practically, the coefficients shown in Fig. 10 are derived from the HD envelopes at 12 200 K, 12 000 K, and 11 800 K by numerical differentiation of the divergence of the convective flux in the $P - T$-plane; the variables $P$ and $T$ are treated as independent. The correct relation between the changes of $P$ and T is fulfilled automatically in the pulsational code as long as the basic assumption of quasistatic changes remains valid.

The lowest one dozen modes of degree $\ell = 1, 2$, and 3 of the HD model at 12 000 K were reanalyzed with the IA2 implementation. In none of the eigenvalues did we find any significant or systematic change compared with IA1 and even with the frozen-in flux approximation. Most important, the IA2 approach did not yield any unstable oscillation modes.

### 3.3. Suppressed shear

Brickhill (1990) showed that the action of turbulent viscosity is able to efficiently suppress any horizontal shear motion within the convective layers, i.e. $d\xi_h/dr = 0$ ($\xi_h$: horizontal displacement) throughout this region. This assumption was explicitly programmed into the linearized equation of mass conservation. The effect of viscous dissipation was neglected. In the pulsation code, the location of the depth down to which $d\xi_h/dr = 0$ applied was varied for test purposes. We found no differences in the results when switching between the innermost mass-shell with positive $\mathbf{F}_C$, the deepest mass-shell in the HD layers with nonvanishing $\mathbf{F}_C$, and the extrapolated depth where $v_{rms} = 10$ cm sec$^{-1}$ (see Freytag et al. 1995).

As mentioned at the end of Sect. 2.2., the values of the convective flux derived from $\nabla_{ad}$ and $\nabla_0$ can differ slightly from the hydrodynamically obtained enthalpy flux. When the convective flux was set equal to the enthalpy flux we found unstable low-order $g$ modes and spherical degrees $\ell = 1, 2$, and 3 with periods below 200 sec (see Fig. 11) in models A4 and A7. We note that the occurrence of instabilities did not depend on the amount of hydrogen floating on the surface of the HD models. The weakest instabilities were found for the $\ell = 1$ modes. With increasing spherical degree the strength of instability increased. The growth rates increased with radial order and changed rapidly to strong damping above a maximum radial order. We restricted the calculations to the potentially observable low-degree modes, the instabilities persist most probably to higher degrees. We found that both, the calculation of convective flux by enthalpy flux *and* vanishing shear were *necessary* ingredients to the oscillation equations for unstable modes to occur.

Furthermore, we also tested the influence of suppressed shear in the Bradley models. Over the spatial domain of nonzero convective flux, we assumed $d\xi_h/dr \equiv 0$. A representative result is included in Fig. 9 as points connected with a dotted line. The high-order modes were substantially stabilized so that the upper period limit for unstable $g$ modes dropped from 1000 sec to roughly 600 sec (for $\ell = 3$ at 12 080 K). Low-order modes (of degree $\ell = 1 - 3$), on the other hand, showed enhanced instability.

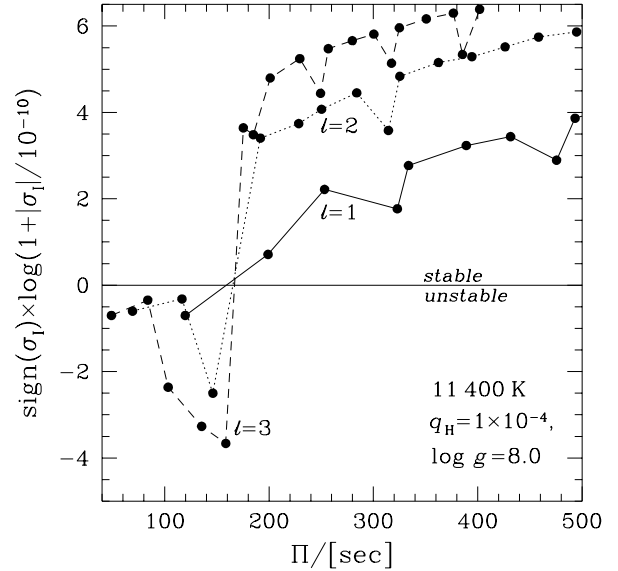

**Fig. 11.** Stability diagram of the lowest-degree $g$ modes for the 11 400 K model A4. When the convective flux was calculated from the enthalpy flux resulting from the hydrodynamical simulations the lowest-order modes of each of the potentially observable spherical degrees $\ell = 1, 2$, and 3 turned unstable.

## 4. Discussion

From our nonadiabatic nonradial oscillation calculations performed on the HD models we found a blue edge at an effective temperature several hundred degrees below what current observational determinations might indicate. If we treat convection with the frozen-in approximation only one unstable $g$ mode is found for the lowest three spherical degrees at $\log g = 8.0$ in the temperature range from 11 400 K to 12 600 K. This unstable mode has, however, nothing to do with the expected ZZ Ceti properties. The kinetic energy-density distribution reveals that it is confined (possibly artificially due to the arbitrariness in the construction of the transition layers) to the deep interior of the white dwarf models. As the outermost layers have very little weight, its imaginary part does not react noticeably on changing the effective temperature of the model.

We realized that the complex eigenvalues of the HD models listed in Tab. 1 did not change noticeably upon implementing instantaneous adaptation of the convective flux in the oscillation equations. The reason might be the shallowness of the convection zone so that it cannot, except for our lowest value of effective temperature, influence significantly the relative pulsational phases and/or the amplitudes of the perturbation quantities. In the formulation of instantaneous adaptation according to Eq. (1) we modified the radial components of the flux perturbation only which dominates the energy exchange despite that the tangential displacement is often several hundred times larger than the radial one in the outermost layers of the DA models.

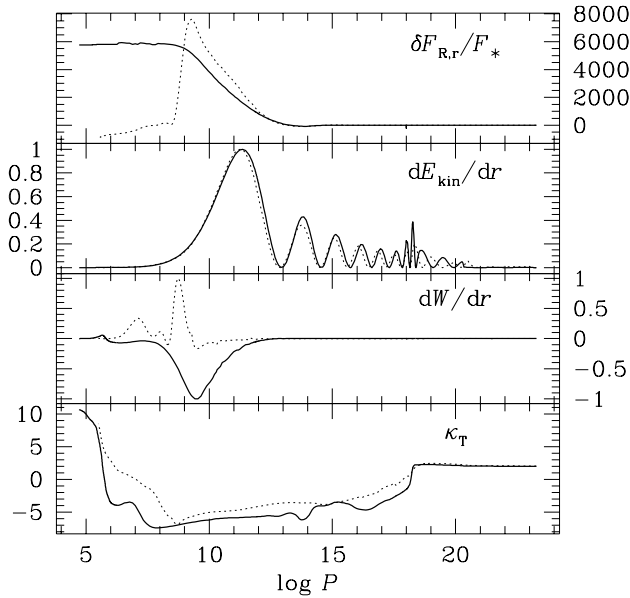

**Fig. 12.** Comparison of eigensolution quantities between the Bradley model at 12 080 K (thin dashed lines) and model C4 (solid lines) for the arbitrarily chosen mode $k = 13$ and $\ell = 3$. Starting at the top, we show the radiative Lagrangian flux perturbation relative to the star's equilibrium surface flux, the differential kinetic energy, the differential work, and the run of the logarithmic opacity derivative at constant temperature $\kappa_T$. The data in two panels in the middle were arbitrarily normalized to their most prominent extrema.

Figure 12 shows a comparison of a selected $g$ mode ($\ell = 3, k = 13$) for DA models around 12 000 K. Bradley's model is pulsationally unstable; properties of this model and mode are shown with dotted lines. The $g$ mode of the HD model C4 on the other hand is pulsationally stable. For both models, the perturbation of the radiative flux becomes roughly frozen-in around $\log P = 8.5$. The thermal time-scales are indeed about the same for both models in these depths. The dynamical weight distribution is surprisingly similar for both models as can be seen in the run of the differential kinetic energy of the modes. Divergences occur mainly around the H/He transitions. These regions were modeled completely independent, after all. Hence, the two modes show a differing trapping behavior in the centermost regions. The $g$ mode in the Bradley model has its main driving peak at the inner edge of the convection zone. Actually, most of the driving in the main peak happens in a region with $\mathrm{d}\kappa_T/\mathrm{d}r < 0$. The $g$ mode of the HD model is – up to an insignificant driving contribution close to the surface – completely damped. In its convection zone, the flux is already frozen-in so that no $\kappa$ effect acts there anymore. The radiative dissipation starts below the convection zone.

The models A4 and A7 were the only ones showing pulsational instabilities; these 11 400 K models have domains of very efficient convective energy transport (see Fig. 13). And only in connection with assumed zero shear across the convection region did these instabilities occur. The local minimum of $F_{\mathrm{rad}}/F_{\mathrm{r}}$ coincides with the peak driving in the work integral. Below the convective flux maximum, the convection zone continues for another three pressure scale heights (including the overshoot region). Hence, in the models A4 and A7 the main driving region can hardly be considered to concur with the base of the convection zone. Nevertheless, we believe that increasing convective efficiency will shift the region of maximum driving towards the bottom of the convection zone.

The comparison models from Bradley, being based on deep efficient convection zones (ML3), showed a large number of unstable oscillation modes not only in the frozen-in flux treatment but also in the vanishing-shear approach. In the latter case the number of unstable modes was reduced, but still, too many unstable modes remained to be in satisfactory agreement with observations.

When adopting instantaneous adaptation of the convective flux and vanishing shear in the convection zone, the unstable low-order $g$ mode disappeared in the stellar models with effective temperatures above 11 800 K. Seemingly, these modes – trapped in the deep interior – could pick up additional small but nevertheless sufficient dissipation in the envelope overcompensating the driving in the C/O core.

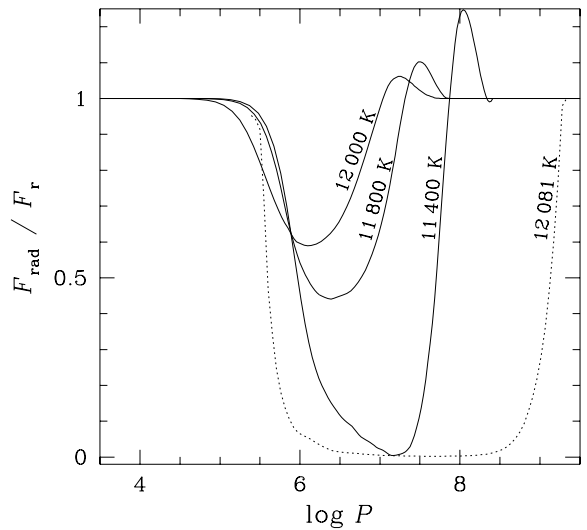

**Fig. 13.** Reduction of the radiative flux in the convection zones of some white-dwarf models. Solid lines are derived from HD models with the appropriate temperatures labeled on the curves. The broken line stands for the 12 080 K model of Bradley with a ML3 convection zone.

The hydrodynamically simulated convection zones of the HD model series are much shallower than what was postulated hitherto in pulsation theory to reproduce instability at effective temperatures derived from spectroscopic analyses of DAV stars (Tassoul et al. 1990, Bradley & Winget 1994, Bergeron et al. 1995). While in the current context the convective efficiencies derived from the HD models appear rather low they are in agreement with otherwise adopted ones. In the surface layers the efficiency from the HD models translates into a corresponding mixing-length parameter of about 1.5 in the ML1 dialect of the MLT. Figure 13 illustrates the discrepancies in the convective efficiencies and the extent of the convection zone for the different models. The solid lines display the results from the hydrodynamic approach. The dashed line shows the outcome from the ML3 treatment in Bradley's 12 081 K model. Evidently, the convection zones of *all* HD models, even the coolest ones, are much shallower than the ML3 solution at 12 081 K. A convective efficiency comparable to the one prevailing in the ML3 model at 12 080 K of Bradley is found only in about 600 K cooler HD models. In contrast to the ML3 approach (of course by the very nature of the formulation) the hydrodynamical convection zones show substantial overshoot regions with negative convective fluxes.

The thermal time-scale of the envelope overlying the driving region is only a few seconds and is hence much shorter than the periods of the unstable modes (between about 80 to 160 sec). Brickhill (1991) argued, however, that even substantial differences between the thermal time-scale of the envelope and the oscillation periods (up to about a factor 25) can still allow for significant driving due to interaction with convection. The 11 400 K HD models could indeed be at the border where such a constructive interaction starts to take effect. In fact, the magnitude of the efficiency of the convective flux appears to be a determining factor to destabilize $g$ modes. This observation applies also to the Bradley models; there, the structure of the convection zone is such that the driving indeed concurs with the base of the convection zone, this location corresponds with the maximum of the convective flux. Furthermore, the main radiative damping in the pulsationally stable HD models is focused on the regions below the convection zone (see Fig. 12). Such a behavior agrees with the picture of a matching of the thermal time-scale of the overlying envelope with the oscillation period of the mode under consideration (Cox 1980).

When building background models we realized that thick hydrogen layers, of the order of $q_H = 10^{-4}$ can lead to a weak hydrogen burning shell close to the H/He transition. At 11 800 K, a H-burning shell achieved a maximum energy generation rate between 6 and 40 erg g$^{-1}$ sec$^{-1}$ depending of particular choices of $\Delta M_H$ and $\Delta M_{He}$. The nonvanishing $\epsilon$ terms (always for equilibrium burning) never noticeably affected the nonadiabatic eigenvalues. In particular, the $\epsilon$ terms never changed the stability property of any mode.

## 5. Summary and concluding remarks

DA white-dwarf models were calculated with the physical properties of the outermost layers deduced from two-dimensional hydrodynamic simulations. Based on these models we performed nonradial, nonadiabatic stability analyses using first canonical assumptions for the treatment of convection. We failed to find any unstable low-order, low-degree oscillation modes over the whole temperature range (which overlaps with present $T_{eff}$ estimates for ZZ Ceti stars) covered by our models (cf. Table 1). The shallow and rather inefficient convection zones – as compared with the typically invoked very efficient convection – in the partial H-ionization region are identified as the reason for our negative result.

Looking at the time scales involved in the hydrodynamically simulated convection zones clearly shows that the frozen-in flux assumption is inappropriate. The thermal time scale in the layers above the driving/dissipation regions (a few seconds), as well as the dynamical turn-over time of an eddy (of the order of a tenth of a second), are much shorter than the oscillation periods of interest. Hence, instantaneous adaptation of convection should approximate reality better.

Only when adopting instantaneous adaptation of convection, and in particular when requiring vanishing shear in the convective layers were we able to find a few pulsationally unstable short-period period – between 80 and 160 sec – $g$ modes in the 11 400 K models. Based on the results from the HD stellar models we conclude that the blue edge of the ZZ Ceti stars should lie between 11 400 and 11 800 K. This is not in agreement with the most recent physical calibrations of observed ZZ Ceti stars. Bergeron et al. (1995) favor a blue edge for these variables around 12 000 K for log $g = 8.0$. Despite recent methodical improvements considerable uncertainty might nevertheless still be present in the spectroscopic analyses.

We agree with Bradley & Winget (1994) and Bergeron et al. (1995) in not finding a dependence on the thickness of the hydrogen layer of the $g$-mode instabilities. If the efficiency of convection is the dominant factor to destabilize some $g$ modes, then we also expect a dependence of the blue edge of the ZZ Ceti instability region on log $g$. Furthermore, as noted in Bergeron et al. (1995), a correlation between $T_{eff}$ and the dominant observed periods were to be expected on theoretical grounds. For cooler stars, the convection zone extends deeper and so does the region of maximum convective flux. The associated thermal time-scale increases and so the interaction with longer periods would be favored.

The HD model envelopes have inner edges of the convective zone extending to $10^{-15}M_*$ (at 12 600 K) and $10^{-14}M_*$ (11 400 K) beneath the surface. These numbers are much in contrast to typical ML3 results; at 11 720 K, the convection zone in Bradley's model extends to about $8 \times 10^{-13}M_*$. An impressive visualization of this difference is provided in Fig. 13. It is not only the shallowness of the convection zone that suppresses sufficient driving for low-order, low-degree $g$ modes in the HD models, but also the accompanying inefficiency of the convective energy transport is responsible for the dilemma. It may appear at first glance that the hydrodynamical simulations have shortcomings that lead to convective zones which are too shallow to drive the pulsations. However, we checked the influence of technical parameters in these calculations, in particular the grid resolution and the size of the computational box, we found no evidence for insufficient modeling. Further, experience gained by other authors in different contexts shows that 2D simulations of convection produce more vigorous flows than the 3D ones (e.g. Muthsam et al. 1995). This behavior is comprehensible since the greater degree of freedom in 3D reduces the stability of the concentrated downdrafts which are the driving entities in the flow. Hence, we expect qualitatively that our hydrodynamical simulations deliver an upper limit for the extent and efficiency of the convective zones in the DAV stars. We conclude that 3D simulations will not remove the discrepancy.

According to our calculations the trapping properties of the $g$ modes do not primarily determine the stability characteristics of those modes defining the blue edge. But we find qualitatively different trapping properties of the oscillation modes in our models that point to a less pronounced mode selection.

The stability analyses of Bradley's models with deep ML3 convective zones gave rise to a large number of unstable $g$ modes in any approximation of the convection perturbation. With his models we worry about finding much too many unstable modes to be compatible with observations.

The work integrals of the unstable modes in the coolest HD models show that pulsational driving always occurs in regions with $d\kappa_T/dr > 0$. This is not necessarily so in Bradley's models; we found unstable modes for which most of the destabilizing work is done in zones with $d\kappa_T/dr < 0$. At least for such cases, a nonstandard driving agent, involving convection must be identified.

We found that turbulence has a marked influence on the structure of the eigenmodes and their excitation. Since we used a rather simple treatment of the effects here, this point deserves further investigation. Despite the not fully grasped role played by convection in exciting the pulsations of ZZ Ceti stars we feel that they belong to the best suited laboratories to study the interplay between pulsation and convection. We finish by provoking that when we find unstable low-degree $g$ modes in DA white-dwarf models by ad hoc tuning of the efficiency of the H-convection zone then it is a fortunate conspiracy only and not the result of a deeper understanding.

*Acknowledgements.* A.G. is grateful for financial support by the Swiss National Science Foundation. B.F. was supported by the Deutsche Forschungsgemeinschaft through grant Ho 596/43-1. B. Paczyński and N.H. Baker are acknowledged for passing on their wisdom to construct simple stellar models and to derive efficient approximations to constitutional relations. P. Bradley kindly provided us with three of his evolutionary white-dwarf models models that enabled us to check the integrity of the pulsation code. H. Saio provided generous advice at different stages, we thank D. Koester for his helpful comments on white dwarf spectroscopy.

## 6. Appendix: Shooting a white dwarf model

The interior structure of the white dwarf models was numerically calculated with a shooting method. A number of simplifying assumptions were to be made. They are discussed, together with a description of the treatment of the constitutive relations, in the following.

The temporally and spatially averaged data from the hydrodynamic simulations of the superficial convective layers of white dwarfs constituted the outermost regions of our spherically symmetric models. The inner edge of the computational box of the simulations provided the outer boundary conditions for the interior solutions. A guess for radius and prescribed values of pressure and temperature at the base of the nonlinear computational grid allowed to start an inward integration to an ad hoc preselected fitpoint, $M_f$, that was chosen between 0.5 and 0.7 $M_\star$. An outward integration started at the center of the star with trial values for central pressure and temperature. At the fitpoint the solution components for pressure, temperature and radius had to join continuously. Continuity of these variables was obtained with a Newton-Raphson iteration of values for the radius of the white dwarf at the outer edge of the shooting region, of the central pressure, and of the central temperature.

The chemical stratification of the stellar models are parameterized in a purely phenomenological way. The surface region

consists of pure hydrogen. The transition to a pure He shell is simulated as:

$$X(q) = \left( \tanh \left( \left( \frac{q - 1 + q_H}{\Delta_H} \right) + 1.0 \right) \right) \cdot \varphi.$$

The transition thickness is determined by $\Delta_H$ and was usually set to be half of $q_H = \Delta M_H/M_\star$, the relative mass depth of the hydrogen blanket. Similarly, $q$ is the relative mass coordinate across the star. The factor $\varphi$ finally allows to construct asymmetric profiles by requiring $X(q) = \min(1, x(q))$, i.e. we can cut off the asymptotic regime of the hyperbolic tangent towards large values of $q$. Physically it means that hydrogen develops an extensive diffusive tail into the helium layer but not vice versa. The same reasoning is applied to obtain the transition from He to the C/O cores of the white dwarf models. Despite this ansatz lacks any physical justification it satisfactorily describes the attempted *qualitative* behavior of the abundance stratification in the models.

Energy generation in the star was assumed to be completely due to thermal cooling of the stellar matter, and it is approximated by

$$\frac{d(L/L_\star)}{d(M/M_\star)} = 1.$$

Full evolution calculations (Tassoul et al. 1990) indicate that such an approach is reasonable for DA variables. These stars have cooled down already to such low luminosities that neutrino emission is not important anymore. For crystallization, on the other hand, the interior of the DAV stars is still not dense enough.

*EOS:* The basic principle of our very simple, but computationally efficient, equation of state is that the total pressure of the stellar matter is computed as the sum of a number of analytically obtainable physical components. For given $\rho$, $T$, and chemical composition the total pressure can be calculated explicitly. Since we use $(P, T, X, Y, C, O, Z)$ as the thermodynamic basis, however, the EOS must be solved iteratively. (The quantity Z is defined in this case as $1 - X - Y - C - O$.) Typically, less than 6 iteration steps reduced the relative change of successive corrections in the pressure below $10^{-6}$.

For a given chemical composition, $\rho$, and $T$ the total pressure is assumed to be composed as follows

$$P = P_{ion} + P_{Coul} + P_e + P_{rad}.$$

The pressure component $P_{ion}$ denotes the ideal, nondegenerate ion pressure

$$P_{ion} = \frac{k}{\mu_{ion} m_u} \rho T.$$

The ionic molecular weight is denoted by $\mu_{ion}$, $m_u$ is the atomic mass unit and $k$ stands for Boltzmann's constant. The non-ideal effect of Coulomb interaction between ions is included in the form given in Tassoul et al. (1990). Radiation pressure is $P_{rad} = aT^4/3$, with $a$ being the radiation constant.

The electron pressure $P_e$ is made up of two components

$$\frac{1}{P_e^2} = \frac{1}{P_{e,nd}^2} + \frac{1}{P_{e,d}^2}.$$

The pressure from nondegenerate electrons, $P_{e,nd}$, is computed by solving the Saha equation for the ionization states of

X, Y, C, and O. The resulting information allows the computation of $\mu_e$ and hence of

$$P_{e,nd} = \frac{k}{\mu_e m_u} \rho T.$$

To avoid discontinuities in physical quantities due to different numerical treatments of the various regions in the $\rho - P$ plane, pressure ionization was included in the Saha equation so that this contribution could be accounted for in the EOS throughout the whole star. Pressure ionization was implemented by adding an additional fictitious potential to the "physical" ionization potentials in the Saha equation. The approach is the same as in Eggleton et al. (1973), the numerical values of the parameters in the pressure-ionization term were adopted from Proffitt & Michaud (1991).

Again, degeneracy pressure of electrons was calculated as a weighted sum

$$\frac{1}{P_{e,d}^2} = \frac{1}{P_{e,dnr}^2} + \frac{1}{P_{e,dr}^2}.$$

The nonrelativistic case is approximated by

$$P_{e,dnr} = 9.91 \times 10^{12} \left(\frac{\rho}{\mu_e}\right)^{5/3},$$

and the relativistic case is

$$P_{e,dr} = 1.23 \times 10^{15} \left(\frac{\rho}{\mu_e}\right)^{4/3}.$$

Our approach to the calculation of the EOS does not ensure thermodynamic consistency. Checks with versions of the Eggleton – Faulkner – Flannery EOS (W. Däppen, private communication) showed that the largest discrepancies occur in the domain of partial electron degeneracy. The relative differences never exceeded 5 %. We continue to use our formulation because of its computational efficiency that is helpful on small computers in particular.

Accounting for a nonvanishing Coulomb interaction in the formulation of the EOS proved crucial to obtain low enough periods – to be compatible with observations – of low-degree oscillation modes in the DA white dwarf models. The magnitude of the density in the stellar models did not change much, the density *gradient* and hence the Brunt-Väisälä frequency reacted sensitively on the inclusion of ion interactions.

*Opacities:* In the pure hydrogen layers of the white dwarf models we used OPAL opacity data. The hydrodynamic simulations were performed with the opacity data used in the ATLAS6 stellar atmosphere program supplemented with contributions from Lyα-features. Most of the simulations were actually performed with frequency-dependent radiation transport. As far as a comparison is possible, the Rosseland opacities derived from the OPAL tables and those used in the HD simulations agree very well. Within an OPAL table we employed bi-rational spline interpolations.

In the very hot and high-density domains of the helium and carbon/oxygen layers the opacities were calculated from the following approximation formulae

$$\frac{1}{\kappa} = \frac{1}{\kappa_{Ross}} + \frac{1}{\kappa_{cond}},$$

where the Rosseland mean $\kappa_{Ross}$ is made up from a "Kramers-opacity" term $\kappa_K$ and an electron-scattering term $\kappa_{es}$

$$\kappa_{Ross} = \kappa_{es} + \kappa_K.$$

The latter two sources of absorption are approximated by

$$\kappa_{es} = \frac{0.2\,(1+X)}{\left[1 + 2.7 \cdot 10^{11}\rho/T^2\right]\left[1 + (T/4.5 \cdot 10^8)^{0.86}\right]},$$

including Thomson and Compton scattering, and

$$\kappa_K = 3.8 \cdot 10^{22}\,(1+X)\,(X + Y + 3\,C + 4\,O)\,\rho T^{-7/2}.$$

The conductive opacity is naïvely estimated as

$$\kappa_{cond} = 2.6 \cdot 10^{-7} \overline{Z} \left(\frac{T}{\rho}\right)^2 \left(1 + \left(\frac{\rho}{2 \cdot 10^6}\right)^{2/3}\right).$$

where we set $\overline{Z} = X + 4\,Y + 8\,Z$.

Analytic approximations to opacity sources in the low temperature domain were not necessary in our applications since detailed opacity tables were used in the pure hydrogen layers of the models, they always extended to temperatures exceeding $10^5$ K; the actual values depended on the assumed mass of the superficial hydrogen blanket.

In the compositional transition region from pure hydrogen to pure helium we interpolated linearly between the OPAL opacity table and the analytic opacities.

Our simplistic treatment of the physics in the interior of the white-dwarf models led to surprisingly accurate models. Comparison with data presented in Tassoul et al. (1990) revealed that the central density deviated about 6% from a model with the same central temperature and with the same mass. For the same luminosity and equal mass, the central temperatures diverged by roughly 20 % whereas our models were hotter.